\documentclass[11pt,letterpaper,prb,amsmath,amsfonts,amssymb,tightenlines]{revtex4}

\usepackage{amsthm}
\usepackage{color}
\usepackage{hyperref}

\newcommand{\tinyspace}{\mspace{1mu}}

\newcommand*{\bra}[1]{\langle #1|}
\newcommand*{\ket}[1]{|#1\rangle}

\newcommand{\tprod}{\otimes}
\newcommand{\smallnorm}[1]{\bigl\lVert\tinyspace#1\tinyspace\bigr\rVert}
\newcommand{\norm}[1]{\left\lVert\tinyspace#1\tinyspace\right\rVert}
\newcommand{\tnorm}[1]{\norm{#1}_{\mathrm{tr}}}
\newcommand{\dnorm}[1]{\norm{#1}_{\diamond}}
\newcommand{\threenorm}[1]{\left\lvert\left\lvert\left\lvert\tinyspace
  #1\tinyspace\right\rvert\right\rvert\right\rvert}
\newcommand{\smthreenorm}[1]{\bigl\lvert\bigl\lvert\bigl\lvert\tinyspace #1\tinyspace\bigr\rvert\bigr\rvert\bigr\rvert}
\newcommand{\abs}[1]{\left\lvert\tinyspace #1 \tinyspace\right\rvert}

\newcommand{\dm}[1]{\dim \mathcal{#1}}
\newcommand{\linear}[1]{\mathbf{L}(\mathcal{#1})}
\newcommand{\density}[1]{\mathbf{D}(\mathcal{#1})}

\newcommand{\transform}[1]{\mathbf{T}(\mathcal{#1})}
\newcommand{\identity}[1]{I_{\mathcal{#1}}}
\newcommand{\nidentity}[1]{\tilde{I}_{\mathcal{#1}}}

\newcommand{\tr}{\operatorname{tr}}

\newcommand{\opv}{\operatorname{\nu}}
\newcommand{\ptr}[1]{\tr_\mathcal{#1}}
\newcommand{\ceil}[1]{\left\lceil #1 \right\rceil}

\newcommand{\class}[1]{\textup{\textrm{#1}}}

\newtheorem{theorem}{Theorem}
\newtheorem{lemma}[theorem]{Lemma}
\newtheorem{proposition}[theorem]{Proposition}
\newtheorem{corollary}[theorem]{Corollary}
\theoremstyle{definition}

\newenvironment{probenv}[1]{\begin{trivlist}\item {\bf #1.}\em}{\end{trivlist}}

\begin{document}


\title{Additivity and Distinguishability of Random Unitary Channels}

\author{Bill Rosgen}
\affiliation{Institute for Quantum Computing and School of Computer Science,
 University of Waterloo, Canada}
\email{wrosgen@iqc.ca}

\begin{abstract}
  A random unitary channel is one that is given by a convex
  combination of unitary channels.  It is shown that the conjectures
  on the additivity of the minimum output entropy and the
  multiplicativity of the maximum output $p$-norm can be equivalently
  restated in terms of random unitary channels.  This is done by
  constructing a random unitary approximation to a general quantum
  channel.  This approximation can be constructed efficiently, and so
  it is also applied to the computational problem of distinguishing quantum
  circuits.  It is shown that the problem of distinguishing random
  unitary circuits is as hard as the problem of
  distinguishing general mixed state circuits, which is complete for
  the class of problems that have quantum interactive proof systems.
\end{abstract}

\maketitle

\section{Introduction}

A quantum channel is \emph{random unitary} if it can be decomposed
into the probabilistic application of one of a finite set of unitary
operations.  More formally, $\Phi$ is random unitary if there exist
unitary operators $U_1, \ldots, U_n$ and a probability distribution
$p_1, \ldots, p_n$ such that
\begin{equation}\label{eqn:rand-unitary}
  \Phi(X) = \sum_{i=1}^n p_i U_i X U_i^*.
\end{equation}
It has been shown by Gregoratti
and Werner~\cite{GregorattiW03} that the random unitary channels
describe exactly the noise processes that can be corrected using
classical information obtained by measuring the environment.  For
channels on qubits the random unitary channels are exactly the unital
channels, but for larger dimensions this is not the
case~\cite{Tregub86,KuemmererM87,LandauS93}.
Audenaert and Scheel have recently
provided necessary and sufficient conditions for a channel to be 
random unitary~\cite{AudenaertS08}.  
Buscemi has also provided an upper bound on the number of unitaries
needed for a random unitary decomposition~\cite{Buscemi06}, as in Equation~\eqref{eqn:rand-unitary}.

A natural question arises from this class of channels: is the
additivity conjecture simplified when restricted to the random unitary
channels?  In the present paper this question is answered in the
negative.
This is done using a method to approximate an arbitrary quantum
channel by a random unitary one.  
A recent survey on the additivity conjecture and a 
few related conjectures that we will also
consider can be found in Ref.~\onlinecite{Holevo06}.
One such conjecture is the
question of the additivity of the minimum output entropy.
The approximation scheme constructed here is also used to show that
this conjecture can be
restricted to the random unitary channels with no loss of generality,  
extending the results of
Fukuda~\cite{Fukuda07} on unital channels.
In addition to these results, this approximation scheme implies
the computational hardness of distinguishing
mixed-state quantum circuits that implement random unitary channels.

All Hilbert spaces considered here are finite dimensional and denoted
by calligraphic letters $\mathcal{H, K, \ldots}$.  The set of all
(bounded) linear operators on a space $\mathcal{H}$ is denoted by $\linear{H}$.
The set of mixed states, or density operators, which are the
positive semidefinite operators with unit trace on the space
$\mathcal{H}$, is denoted $\density{H}$.  The set $\density{H}$ is
compact and convex.  The
extreme points of $\density{H}$ are called pure states, which are given by the
rank-one projectors $\ket \psi \bra \psi$ onto states of $\mathcal{H}$.
The notation
$\transform{H,K}$ is used for the set of admissible maps from
$\linear{H}$ to $\linear{K}$.  An admissible map,
hereafter be called a channel, is one that is 
completely positive and trace preserving.
The notation $\nidentity{H}$ will be used to denote the maximally
mixed state on the space $\mathcal{H}$, i.e. $\nidentity{H} = \identity{H} /
\dim \mathcal{H}$.

The entropy of a quantum state $\rho$ is given by $S(\rho) = - \tr
\rho \log \rho$.  This quantity can be seen as a measure of purity, as
ranges between $S(\ket \psi \bra \psi) = 0$ for any pure state and
$S(\nidentity{H}) = \log \dim \mathcal{H}$ for the maximally mixed state.
A description of many of the fundamental 
properties of the von Neumann entropy can be found
in Ref.~\onlinecite{NielsenC00}.  
Of particular significance here is the concavity, which is given by
$ S( \sum_i p_i \rho_i) \geq \sum_i p_i S(\rho_i). $

The classical capacity of a single use of a channel $\Phi$ is given by
the $\chi$-capacity~\cite{Holevo98, SchumacherW97}
\[ C_\chi(\Phi) = \max \big[ S( \Phi( \sum_i p_i \rho_i ) ) - \sum_i p_i
  S( \Phi(\rho_i)) \big], \]
where the maximum is taken over all convex mixtures $\sum_i p_i
\rho_i$ of quantum states.  This quantity is also referred to as the
``one-shot'' or ``one-step'' capacity of $\Phi$.  A central question
in quantum information theory is whether is quantity is
\emph{additive}, i.e. does entangling inputs across multiple uses of
the channel increase the capacity?  This question was first raised
in Ref.~\onlinecite{BennettF+97}, and the standing conjecture is that
\begin{equation}\label{eqn:chi-additivity}
  C_\chi(\Phi \tprod \Psi) \stackrel{?}{=} C_\chi(\Phi) + C_\chi(\Psi),
\end{equation}
which is the statement that entangled inputs do not increase the
classical information carrying capacity of quantum channels.

Closely related to the additivity of the $\chi$-capacity is the
question of the additivity of the minimum output entropy,
defined by $S_{\min}(\Phi) = \min_\rho S(\Phi(\rho))$, where the
minimization is over all density operators.  The additivity of this
quantity, given by
\begin{equation}\label{eqn:smin-additivity}
  S_{\min}(\Phi \tprod \Psi) \stackrel{?}{=} S_{\min}(\Phi) + S_{\min}(\Psi),
\end{equation}
was first studied by King and Ruskai~\cite{KingR01}, who attribute this
conjecture to Shor.
This conjecture is connected to the additivity of the $\chi$-capacity by
a result of Shor~\cite{Shor04} that shows that both of these
conjectures are equivalent to a third conjecture:
the strong superadditivity of the entanglement of formation.

One potential path to resolving these conjectures lies in yet another
conjecture.  This is the conjectured multiplicativity of the maximum
output $p$-norm, first stated by Amosov, Holevo, and
Werner~\cite{AmosovH+00}.  This conjecture involves the
maximum output $p$-norm of a quantum channel
$\Phi \in \transform{H,K}$.  This quantity, for $p \in [1,\infty)$ is
given by
\[ \opv_p(\Phi)  
  = \max_{\rho \in \density{H}} \norm{\Phi(\rho)}_p
  = \max_{\rho \in \density{H}}
  \left( \tr \abs{ \Phi(\rho) }^p \right)^{\frac{1}{p}}, \]
which is simply the $p$-norm of the singular values of $\Phi(\rho)$,
maximized over all inputs $\rho$.
This is extended to the case of $p = \infty$ in the
usual way by replacing the sum in the $p$-norm 
with a maximization over the singular
values of $\Phi(\rho)$.
The conjecture of Amosov, Holevo, and Werner~\cite{AmosovH+00}
corresponding to this quantity is that it is multiplicative with
respect to the tensor product of two channels, i.e. that
\begin{equation}\label{eqn:pnorm-mult}
  \opv_p(\Phi \tprod \Psi) \stackrel{?}{=} \opv_p(\Phi) \opv_p(\Psi).
\end{equation}
This conjecture implies the additivity of the minimum output entropy,
given in Equation~\eqref{eqn:smin-additivity}, as the derivative of
$\opv_p(\Phi)$ for $p$ approaching one gives an expression for
$S_{\min}(\Phi)$.
The multiplicativity conjecture is known to fail for any fixed
$p>1$~\cite{HaydenW08}.
This does not eliminate
interest in this quantity, however, as the conjecture can be weakened
to ask if, for given $\Phi, \Psi$, does there exist a
sequence $\{p_n\}$ converging to one with $p_n > 1$ for which
Equation~\eqref{eqn:pnorm-mult} holds?
This weakened conjecture still implies the additivity conjectures
given by Equations~\eqref{eqn:chi-additivity} and~\eqref{eqn:smin-additivity}.

There are two main approaches to resolving these conjectures.
The first of these is to consider restricted classes of channels for
which the conjectures can be shown true, in the hope that such a
strategy will yield some insight into the general problem.
The entanglement breaking
channels are one such class, with the additivity of the minimum output
entropy shown by Shor~\cite{Shor02} and the multiplicativity of the
$p$-norm shown by King~\cite{King03}.  This class of channels contains
many important channels, such as the completely depolarizing channel.
Another class of channels for which the additivity and
multiplicativity conjectures are known to hold is the class of unital
channels on qubits~\cite{King02}, which is particularly interesting in
the context of this paper, as the unital qubit channels are exactly
the random unitary channels on qubits.

The second approach taken toward resolving these conjectures is to show
that they remain equivalent when restricted to certain classes of
channels, in the hope that the restrictions will aid search for either
a proof or a counterexample.  
One such equivalence, due to Fukuda~\cite{Fukuda07}, shows that the
conjectures on the additivity of the minimum output entropy and the
multiplicativity of the $p$-norm lose no generality when they are
restricted to unital channels, which are the channels $\Phi \in
\transform{H,H}$ satisfying $\Phi(\identity{H}) = \identity{H}$.
The same approach is taken by Fukuda and Wolf~\cite{FukudaW07}, who
show that, with no loss in generality, these conjectures can be
restricted to two copies of the same channel, i.e. letting $\Psi =
\Phi$ in Equations~\eqref{eqn:smin-additivity} and~\eqref{eqn:pnorm-mult}.

In the present paper we take the second approach, extending the
results of Fukuda~\cite{Fukuda07} on unital channels to the random
unitary case.  This is done by constructing a random unitary
approximation to an arbitrary channel in Section~\ref{scn:channels},
and showing that this approximation is not too far from the original
channel in Section~\ref{scn:properties}.  In
Section~\ref{scn:mult} it is shown how this approximation can be used
to restrict the multiplicativity conjecture to the random unitary
case.  
This is of less interest than the same result for additivity conjecture on the
minimum output entropy, which appears in Section~\ref{scn:add},
but the argument for the $p$-norm case is presented first, as it is both
similar to and simpler than the argument for the minimum output entropy.
Finally, the paper concludes with a discussion of a circuit
implementation of the approximation scheme, in
Section~\ref{scn:circuits}, which is used in
Section~\ref{scn:distinguish} to show that the computational problem
of distinguishing two mixed state quantum circuits is made no easier
by adding the
restriction that the circuits implement random unitary transformations.

\section{Random Unitary Approximation}\label{scn:channels}

Stinespring's Dilation Theorem~\cite{Stinespring55} states that any
quantum channel $\Phi$ can be written as
\begin{equation}\label{eqn:stinespring}
  \Phi(X) = \ptr{B} U (\ket 0 \bra 0 \tprod X) U^*,
\end{equation}
for $U$ a unitary operation.  There are two operations in this
representation that are not random unitary, as defined by
Equation~\eqref{eqn:rand-unitary}.  These operations are the partial
trace over the system $\mathcal{B}$, and the introduction of the
ancillary system in the state $\ket 0$.  To find an
approximation to $\Phi$ that is random unitary, we will need to deal
with both of these operations.

Fixing notation, let $\Phi$ be a completely positive and trace
preserving map from $\linear{H}$ to $\linear{K}$.  Representing
$\Phi$ as in Equation~\ref{eqn:stinespring}, let $\mathcal{A}$ be the
space containing the ancillary space starting in the $\ket 0$ state, and let
$\mathcal{B}$ be the space that is traced out.  This implies that $U$
is a unitary map from $\mathcal{A \tprod H}$ to $\mathcal{K \tprod B}$.

To avoid tracing out the system in the space $\mathcal{B}$ the partial trace
may be replaced by the operation $N_\mathcal{B}$ that takes
the state in $\mathcal{B}$ to the completely mixed state.  This operation
can be implemented as a random unitary operation as the uniform
mixture of the discrete Weyl
operators)~\cite{AmbainisM+00,BoykinR03,HaydenL+04}.
The discrete Weyl operators are also known as the generalized Pauli
operators, as they are one generalization of the Pauli $X$ and $Z$
operators to higher dimensional systems.
It is not difficult to see that for
$\rho$ a density matrix on $\mathcal{K \tprod B}$,
\begin{equation}\label{eqn:traceout}
   N_{\mathcal{B}}(\rho) = \left( \ptr{B} \rho \right) \tprod \nidentity{B}.
\end{equation}
This implies that if the system to be traced out instead has
$N_\mathcal{B}$ applied to it, the resulting state is the same, up to
a tensor factor of a maximally mixed state in the space $\mathcal{B}$.  
This factor will
change both the minimum output entropy and the maximum output $p$-norm
by a fixed value that will not affect the additivity or
multiplicativity of these quantities.

Replacing the introduction of the ancillary space $\mathcal{A}$ with a
random unitary operation is more complicated.  The strategy employed
is to expand the input of the transformation to include the space
$\mathcal{A}$.  The input state of this system will not, in general,
be the desired state $\ket 0$, and so an additional operation is
needed to force this to be the case for any input that minimizes the
output entropy.  As we are only interested in the minimum output
entropy and the maximum output $p$-norm, those inputs on which the
resulting channel produces an output with high entropy can be ignored,
as they will be far from those inputs that achieve the minimum
(resp. maximum).

To this end, the ideal operation to perform this forcing does not
alter any input state of the form $\ket 0 \bra 0 \tprod \sigma$, but
takes any orthogonal state to the completely mixed state $\nidentity{A
  \tprod H}$.  This operation is, unfortunately, not random unitary,
as it is not unital.  
A closely related strategy that is random unitary is to project the
input state either onto the subspace $S_0 = \ket 0 \tprod \mathcal{H}$
or the orthogonal subspace $S_0^\perp = \ket{0}^\perp \tprod
\mathcal{H}$.  This projection is then be followed by a mixing
operation on the subspace $S_0^\perp$.
This mixing process is be introduced first.  It is given by
the channel $M$ that does not affect
the subspace $S_0$ but completely mixes $S_0^\perp$.
More concretely, on a state $\rho = q \rho_{S_0} + (1-q)
\rho_{S_0^\perp}$ where $\rho_{S_0} = \ket 0 \bra 0 \tprod \sigma$ is
a density operator on $S_0$ and $\rho_{S_0^\perp}$ a
density operator on $S_0^\perp$, the output of $M$ is given by
\begin{align}
  M(\rho) = q M(\rho_{S_0}) + (1-q) M(\rho_{S_0^\perp})
          &= q \rho_{S_0} + (1-q) \tilde{I}_{S_0^\perp} \nonumber \\
          &= q \ket 0 \bra 0 \tprod \sigma
            + (1-q) \frac{\identity{A} - \ket 0 \bra 0}{\dim \mathcal{A}-1}
              \tprod \nidentity{H}. \label{eqn:mixing-approx}
\end{align}
The channel $M$ can be implemented as a random unitary channel in the
same way as the completely depolarizing channel: a uniform mixture of
the discrete Weyl operators, except here these operators are taken
over the subspace $S_0^\perp$.

This channel is not exactly the desired one.  If the output of $M$ on
$\rho_{S_0^\perp}$ in Equation~\eqref{eqn:mixing-approx} were the
completely mixed state on $\mathcal{A \tprod H}$ and not the subspace 
$S_0^\perp$ then this process would create an essentially
error-free random unitary approximation of the original channel (for
the purpose of minimizing the output entropy).  Fortunately, the error
involved at this step will be shown, in
Lemma~\ref{lem:ancilla-mixing}, to be $O(1 / \dim{\mathcal{A}})$,
and so by taking the space $\mathcal{A}$ large enough we will be able
to approximate the ideal case.

There is one further convenient property that this mixing channel does
not satisfy: it does not remove coherences between the subspaces $S_0$
and $S_0^\perp$.
If the channel $M$ had this property, then an
equation similar to Equation~\eqref{eqn:mixing-approx} would hold for all
input states $\rho$, not just those states that 
have no entanglement between the subspaces $S_0$ and $S_0^\perp$.
This property will
be essential to the analysis that follows, and so the additional
operation that decoheres these two subspaces needs to be applied
before the mixing operation $M$.  This operation, $D$, can be
implemented by leaving the state unchanged with probability one half
and applying a unitary $U$ with probability one half, where the
action of $U$ on basis states is given by
\begin{align*}
  U \ket i &= \begin{cases}
    \ket i  & \text{if $\ket i \in S_0$},\\
    -\ket i & \text{if $\ket i \in S_0^\perp$}.
  \end{cases}
\end{align*}
In other words, $U$ applies a phase of $-1$ to states in
$S_0^\perp$ and does not change states in $S_0$.  When $U$ is applied
with probability one half the result is complete dephasing between the
two subspaces.  This can be seen by observing that this is the
restriction of the completely dephasing channel, as considered
in Ref.~\onlinecite{DattaF+06}, to a system with only two orthogonal states,
which are here given by the subspaces $S_0$ and $S_0^\perp$.
Alternately, when this is applied to a density matrix expressed in the
computational basis, the result is, by a simple calculation, the
zeroing of the off-diagonal elements of the first row and column.
When this operation, $D$, is applied to a density operator $\rho \in
\density{A \tprod H}$, the result is
\begin{align}\label{eqn:decoherence}
  D(\rho) = q \rho_{S_0} + (1-q) \rho_{S_0^\perp}
          = q \ket 0 \bra 0 \tprod \sigma + (1-q) \rho_{S_0^\perp},
\end{align}
where $\rho_{S_0} = \ket 0 \bra 0 \tprod \sigma$ is a density operator on
the subspace $S_0 = \ket 0 \tprod H$, $\rho_{S_0^\perp}$ is a density
operator on $S_0^\perp$, and $0 \leq q \leq 1$.

Combining Equations~\eqref{eqn:mixing-approx}
and~\eqref{eqn:decoherence}, the output of $D$ followed by $M$ on a
density operator $\rho$ on $\mathcal{A \tprod H}$ is given by a state
of the form
\begin{align*}
  (M \circ D)( \rho ) &= q M(\ket 0 \bra 0 \tprod \sigma) + (1-q) M(\rho_{S_0^\perp})
                  = q \ket 0 \bra 0 \tprod \sigma
            + (1-q) \frac{\identity{A} - \ket 0 \bra 0}{\dim \mathcal{A}-1}
              \tprod \nidentity{H}.
\end{align*}
This operation $M \circ D$ will be used as a way to force any input
that results in a low output entropy to be close to the subspace $S_0$
of inputs having the `ancilla' space $\mathcal{A}$ in the desired
$\ket 0$ state.  On these inputs the constructed random unitary
channel will behave in a similar way to the original channel that is
being approximated.  On inputs that are far from this subspace, the
resulting state has high entropy, and so it will not be close to a
state minimizing the output entropy.
As $M \circ D$ mixes the input, conditional on
the state being in the subspace $S_0^\perp$, this operation will be
referred to as the conditional mixing procedure. 

Putting all of these pieces together, given a channel $\Phi(\rho) =
\ptr{B} U (\rho \tprod \ket 0 \bra 0) U^*$, the random unitary
approximation $\Phi'$ is constructed as
\begin{equation}\label{eqn:construction}
  \Phi'(\rho) =   
  N_\mathcal{B} \left( 
    U \left[ 
      (M \circ D)(\rho) 
    \right] U^* 
  \right),
\end{equation}
which, more plainly, is simply the application of the conditional mixing
procedure, the unitary operation from a Stinespring
dilation of $\Phi$, and finally the completely mixing channel to the
space that would have been traced out by $\Phi$.  As the composition
of random unitary transformations remains random unitary, the channel
$\Phi'$ will be a random unitary channel.

It will be useful to observe that the channel $\Phi'$ specified in
Equation~\eqref{eqn:construction} can be used to simulate the channel
$\Phi$.  This occurs when the input $\ket 0 \bra 0 \tprod \sigma$,
i.e. an input in the space $S_0$, is provided to $\Phi'$.
This is argued in the following proposition.

\begin{proposition}\label{prop:simulation}
  Let $\Phi$ be a quantum channel from $\linear{H}$ to $\linear{K}$.
  If $\Phi'$ is the random unitary channel mapping $\linear{A \tprod H}$
  to $\linear{K \tprod B}$ that is constructed from $\Phi$ in
  Equation~\eqref{eqn:construction}, then
  \[ \Phi'(\ket 0 \bra 0 \tprod \sigma) = \Phi(\sigma) \tprod \nidentity{B}. \]  

  \begin{proof}
    Notice that both $D$ and $M$ do not
    affect this input: the decoherence operation $D$ does not affect
    the state as it is in the subspace $S_0$ and 
    $M$ does not affect the state by
    Equation~\eqref{eqn:mixing-approx}.  Thus, the output of the
    channel $\Phi'$ is
    \begin{align*}
      \Phi'(\ket 0 \bra 0 \tprod \sigma) 
      &= N_\mathcal{B} \left( 
          U \left[ 
            (M \circ D)(\ket 0 \bra 0 \tprod \sigma) 
          \right] U^* 
        \right)  \\
      &= N_\mathcal{B} \left( 
        U (\ket 0 \bra 0 \tprod \sigma) U^* 
        \right)  \\
      &= \ptr{B} \left( 
        U (\ket 0 \bra 0 \tprod \sigma) U^* 
        \right) \tprod \nidentity{B} \\
      &= \Phi(\sigma) \tprod \nidentity{B},
    \end{align*}
    where the penultimate equality is an application of
    Equation~\eqref{eqn:traceout}. 
  \end{proof}
\end{proposition}

Combining this proposition with Equation~\eqref{eqn:decoherence}
that demonstrates the effect of the $M \circ D$ on states not of this
form, and the observation that applying $M \circ D$
twice has no further effect than applying it once, the output of $\Phi'$
on an arbitrary input state $\rho$ is given by
\begin{equation}\label{eqn:output}
  \Phi'(\rho)
  = p \Phi'(\ket 0 \bra 0 \tprod \sigma) + (1-p) \Phi'(\rho_{S_0^\perp})
  = p \Phi(\sigma) \tprod \nidentity{B}  + (1-p) \Phi'(\rho_{S_0^\perp}),
\end{equation}
where as in Equation~\eqref{eqn:decoherence} $\rho_{S_0^\perp}$ is a density
operator on the subspace $S_0^\perp$ of inputs orthogonal to those
with the state $\ket 0$ on the space $\mathcal{A}$.
The major technical portion of the results that follow lies in
bounding the distance from the maximally mixed state of the second term in
this equation, from which most of the results follow.

\section{Properties of the Constructed Channel}\label{scn:properties}

In this section some basic results on the random unitary approximation
of a channel are shown.  Throughout this
section, and the following two sections
$\Phi$ will represent the original transformation and $\Phi'$ will
represent the random unitary transformation constructed from it as in
Equation~\eqref{eqn:construction}.

As a first step to showing that $\Phi'$ approximates $\Phi$ it is shown
that random unitary transformations cannot increase the distance of a
state from the completely mixed state.  This lemma shows that
the output of a random unitary transformation cannot be more pure than the
input.  The extra space $\mathcal{B}$ appearing in this lemma will
correspond to a reference system needed for the results in 
Section~\ref{scn:distinguish}.

\begin{lemma}\label{lem:ru-dist-noise}
  Let $\threenorm{\cdot}$ be a unitarily invariant norm on $\linear{A
  \tprod B}$.  If $\Psi \in \transform{A, A}$ is random unitary,
  then for any $\rho \in \density{A \tprod B}$
  \[ \smthreenorm{ (\Psi \tprod \identity{B})(\rho) 
             - \nidentity{A} \tprod \ptr{A} \rho}
  \leq \smthreenorm{ \rho
              - \nidentity{A} \tprod \ptr{A} \rho} \]
  \begin{proof}
    As $\Psi$ is random unitary, let $\Psi(X) = \sum_i p_i U_i X
    U_i^*$ with the $U_i$ unitary, $0 \leq p_i \leq 1$, and $\sum_i
    p_i = 1$.  Using this decomposition
    \begin{align*}
      \smthreenorm{ (\Psi \tprod \identity{B})(\rho) 
             -  \nidentity{A} \tprod \ptr{A} \rho}
      &= \smthreenorm{ \sum_i p_i (U_i \tprod I) \rho (U_i^* \tprod I)
             -  \nidentity{A} \tprod \ptr{A} \rho} \nonumber \\
       &\leq \sum_i p_i \smthreenorm{ (U_i \tprod I) \rho (U_i^* \tprod I)
              -  \nidentity{A} \tprod \ptr{A} \rho}.
    \end{align*}
    Using the fact that $U_i \nidentity{A} U_i^* = \nidentity{A}$, 
    and the unitary invariance of the norm, this becomes
    \begin{align*}
      \sum_i p_i \smthreenorm{ (U_i \tprod I) ( \rho -  \nidentity{A} \tprod
        \ptr{A} \rho) (U_i^* \tprod I)}
      &= \sum_i p_i \smthreenorm{\rho -  \nidentity{A} \tprod \ptr{A} \rho}\\
      &=  \smthreenorm{ \rho
             -  \nidentity{A} \tprod \ptr{A} \rho}.
    \end{align*}
    Combining these equations yields the statement of the lemma.     
  \end{proof}
\end{lemma}

This lemma can be used to show not only that the conditional mixing procedure
sends states in the subspace $S_0^\perp$ of states where
the ancillary space is not in the $\ket 0$ state
to states that are almost completely mixed, but that the channel $\Phi'$
also has this behaviour.  Before doing this, however, the Lemma is
extended to the case of the von Neumann entropy, where the proof is essentially
identical, with the exception that the triangle inequality is replaced
by concavity.

\begin{corollary}\label{cor:ru-inc-entropy}
  If $\Psi \in \transform{A,A}$ is random unitary, and $\rho \in
  \density{A}$, then
  $ S(\rho) \leq S(\Psi(\rho)). $

  \begin{proof}
    Let $\Psi(\rho) = \sum_i p_i U_i \rho U_i^*$ as in
    Lemma~\ref{lem:ru-dist-noise}.
    Using this
    notation, and the concavity of the von Neumann entropy
    \begin{align*}
      S(\Psi(\rho)) 
      = S(\sum_i p_i U_i \rho U_i^*)
      \geq \sum_i p_i S(U_i \rho U_i^*)
      = \sum_i p_i S(\rho)
      = S(\rho),
    \end{align*}
    where the unitary invariance of the von Neumann entropy has been
    used in the penultimate equality.    
  \end{proof}
\end{corollary}

The next lemma shows that when the input is in the subspace $S_0^\perp$
the output of $\Phi'$ is
very close to completely mixed.  The distance measure used in the
lemma is the trace norm, but this can be applied to the case of the
maximum output $p$-norm due to the fact that $\tnorm{\rho} =
\norm{\rho}_1 \geq \norm{\rho}_p$ for all $p \in [1,\infty]$.
This lemma forms a significant
portion of the proof of the main results on the additivity and
multiplicativity conjectures.

\begin{lemma}\label{lem:ancilla-mixing}
  On input states $\rho \in S_0^\perp$ the output of $\Phi'$ satisfies
  \[ 
     \tnorm{\Phi'(\rho) - 
       \nidentity{A \tprod H}}
     \leq \frac{2}{\dim \mathcal{A}}.\]

  \begin{proof}
    On input $\rho \in S_0^\perp$
    the operation $D$ that introduces decoherence between $S_0$ and
    $S_0^\perp$has no effect.  This implies that the output of $M
    \circ D$ on $\rho$ is given by setting $q = 0$ in
    Equation~\eqref{eqn:mixing-approx}, which is
    \begin{equation}\label{eqn:state-after-mixing}
      \frac{1}{\dim \mathcal{A} - 1} 
       \left( \identity{A} - \ket 0 \bra 0 \right)
       \tprod \nidentity{H},
    \end{equation}
    Setting $d = \dim \mathcal{A}$, we can then compute the distance
    from the completely mixed state as
    \begin{equation}
      \tnorm{ \frac{\identity{A} - \ket 0 \bra 0}{d - 1} 
              \tprod \nidentity{H}
            -  \frac{\identity{A}}{d} 
              \tprod \nidentity{H} }
      = \tnorm{ \frac{\identity{A} - d \ket 0 \bra 0}{d(d-1)}}
      \leq \frac{d-1}{d(d-1)} + \frac{d-1}{d(d-1)}
      = \frac{2}{d}. \label{eqn:norm-after-mixing}
    \end{equation}
    Finally, by noting that the remainder of the transformation
    $\Phi'$ is random unitary, an application of Lemma~\ref{lem:ru-dist-noise}
    yields the desired bound.    
  \end{proof}
\end{lemma}

Once again we can extend this result to the case of the von Neumann
entropy.  In this case we do not simply repeat the same method of
proof, but instead extend the result to the entropy using the
relationship between the trace distance and the entropy given by
Fannes' inequality.
This extension requires that $\dm{A} \geq \dm{H}$, but this can be
assured by considering only those dilations of $\Phi$ with this
property.  In the following corollary we set $m = \log \dm{A}$ for
convenience, but no assumption is made that this value is an integer.
\begin{corollary}\label{cor:mixing-entropy}
  Let $m = \log \dm{A}$.  If $\dm{A} \geq \dm{H}$, $m \geq 3$, 
  and $\rho \in S_0^\perp$, then
  \[ S(\Phi'(\rho)) 
     \geq S(\nidentity{A \tprod H}) - \frac{m}{2^{m-3}}. \]

  \begin{proof}
    Let $\hat\rho$ be the state in
    Equation~\eqref{eqn:state-after-mixing} of the proof of
    Lemma~\ref{lem:ancilla-mixing}.  This is the state after the
    conditional mixing procedure of $\Phi'$ has been applied to the
    input.  For convenience, set $\delta = \smallnorm{\hat{\rho} - \nidentity{A \tprod H}}_{\tr}$.
    By Equation~\eqref{eqn:norm-after-mixing}, this trace distance
    between satisfies $\delta \leq 2/\dm{A} = 2^{-(m-1)}$. Applying Fannes'
    inequality~\cite{Fannes73} (see also Ref.~\onlinecite{NielsenC00})
    yields
    \begin{align*}
      \abs{ S(\hat\rho) - S(\nidentity{A \tprod H})}
      &\leq (\log \dm{H \tprod A}) \delta - \delta \log \delta \\
      &\leq \frac{\log \dm{H}+m}{2^{m-1}} + \frac{m}{2^{m-1}} \\
      &\leq \frac{m}{2^{m-3}},
    \end{align*}
    where the first inequality is by Fannes' inequality, 
    and the second inequality follows from fact that $-x \log x$ is
    monotone for $x \in [0, 1/e]$, 
    and $\delta \leq 2^{-(m-1)} < 1/e$ whenever $m \geq 3$.

    By Corollary~\ref{cor:ru-inc-entropy} applying the remainder of
    $\Phi'$ to the state $\hat\rho$ cannot decrease the entropy, as
    this portion of $\Phi'$
    is random unitary, and so the previous equation implies that
    \[ S(\Phi'(\rho))
       \geq S(\hat\rho)
       \geq S(\nidentity{A \tprod H}) - \frac{m}{2^{m-3}}, \]
    as in the statement of the lemma.    
  \end{proof}
\end{corollary}

\section{Multiplicativity of Random Unitary Transformations}\label{scn:mult}

In this section the construction of Section~\ref{scn:channels} is used
to show some results about the multiplicativity of the maximum output
$p$-norm and random unitary channels.  The main result is that, for $p
< \infty$, the
$p$-norm of the tensor product of two channels is multiplicative if
and only if the $p$-norm is multiplicative on the random unitary
approximations to these channels.

As a first step towards this theorem, it is shown that the random unitary
channel $\Phi'$ constructed from $\Phi$ in
Equation~\eqref{eqn:construction} is a good approximation with
respect to the $p$-norm.

\begin{theorem}\label{thm:approx}
  If $\Phi \in \transform{H,K}$, then the random unitary
  $\Phi' \in \transform{A \tprod H, K \tprod B}$
  satisfies
  \[ \opv_p (\Phi)
     \leq \frac{\opv_p(\Phi')}{\smallnorm{\nidentity{B}}_p}
     \leq \opv_p(\Phi) + \frac{2 \dm{B}}{\dm{A}}. \]
  \begin{proof}
    For convenience, let $d = \dm{A}$.
    The first inequality is simple:
    $\Phi'(\ket 0 \bra 0 \tprod \rho) = \Phi(\rho) \tprod
    \nidentity{B}$ by Proposition~\ref{prop:simulation}, and so it is clear
    that $\opv_p (\Phi) \smallnorm{\nidentity{B}}_p \leq \opv_p(\Phi')$,
    by the multiplicativity of $\norm{\cdot}_p$ with respect to the
    tensor product of states.

    To prove the second inequality let $\rho \in \density{A \tprod H}$
    be a state such that
    \[ \opv_p(\Phi') = \norm{\Phi'(\rho)}_p. \]
    Such a state exists by the compactness of $\density{A \tprod H}$.
    The output of $\Phi'$ on $\rho$ is given by
    Equation~\eqref{eqn:output}, applying the triangle inequality to
    this yields
      \begin{equation*}
       \smallnorm{\Phi'(\rho)}_p
       = \smallnorm{q \Phi(\sigma) \tprod \nidentity{B}  
          + (1-q) \Phi'(\rho_{S_0^\perp})}_p  
       \leq q \smallnorm{\Phi(\sigma) \tprod \nidentity{B}}_p
          + (1-q) \smallnorm{\Phi'(\rho_{S_0^\perp})}_p.  
    \end{equation*}
    Applying Lemma~\ref{lem:ancilla-mixing} to this gives
    \begin{equation*}
       \smallnorm{\Phi'(\rho)}_p
       \leq q \smallnorm{\Phi(\sigma) \tprod \nidentity{B}}_p
           + (1-q)
           \left( \smallnorm{\nidentity{K \tprod B}}_p + \frac{2}{d} \right).
    \end{equation*}
    Then, as the norm $\norm{\cdot}_p$ is multiplicative with respect
    to the tensor product of states, and $\smallnorm{\nidentity{K}}_p \leq
    \norm{\xi}_p$ for any state $\xi \in \density{K}$,
    \begin{equation*}
      \norm{\Phi'(\rho)}_p
      \leq    q \smallnorm{\Phi(\sigma)}_p \smallnorm{\nidentity{B}}_p
           + (1 - q) \left( \smallnorm{\nidentity{K}}_p
           \smallnorm{\nidentity{B}}_p + \frac{2}{d} \right)
      \leq \smallnorm{\Phi(\sigma)}_p \smallnorm{\nidentity{B}}_p
           + \frac{2}{d}.
    \end{equation*}
    Finally, by the choice of the input $\rho$
    \begin{equation*}
      \opv_p(\Phi') = \norm{\Phi'(\rho)}_p
      \leq \opv_p(\Phi) \smallnorm{\nidentity{B}}_p 
        + \frac{2}{d},
    \end{equation*}
    which completes the proof of the theorem, as
    $\smallnorm{\nidentity{B}}_p = \dm{B}^{1/p-1}$.
  \end{proof}
\end{theorem}

With this approximation result, the main theorem on the maximum output
$p$-norm can be shown.  This extends part of the work done by
Fukuda~\cite{Fukuda07} on unital channels to the random unitary case.

\begin{theorem}\label{thm:pnorm-mult}
  If $\Phi, \Psi \in \transform{H,K}$ and $p \in [1,\infty)$, then 
  \[ \opv_p(\Phi \tprod \Psi)  = \opv_p(\Phi) \opv_p(\Psi) \]
  if
  \[ \opv_p(\Phi'_d \tprod \Psi)  = \opv_p(\Phi'_d) \opv_p(\Psi), \] 
  for all sufficiently large $d$, where $\Phi'_d$ is the random unitary
  extension of the channel $\Phi$ obtained by applying the construction of
  Section~\ref{scn:channels} to a Stinespring dilation of $\Phi$ using
  a $d$-dimensional ancillary space.

  \begin{proof}
    As adding ancillary space to $\Phi'$ increases both $\dm{A}$ and
    $\dm{B}$, by taking $d = \dm{A}$ large enough is can be assumed
    that $\dm{B} \leq 2d$.
    Let $\epsilon > 0$, and choose $d$ so that
    $2 \dm{B}^{1 - 1/p} / d \leq 2 / d^{1/p} < \epsilon$.
    Then, as $\Phi'_d(\ket 0 \bra 0 \tprod \rho) =
    \Phi(\rho) \tprod \nidentity{B}$ by Proposition~\ref{prop:simulation},
    \begin{align*}
      \opv_p(\Phi \tprod \Psi)
      &\leq \frac{\opv_p(\Phi'_d \tprod \Psi)}{\smallnorm{\nidentity{B}}_p}.
    \end{align*}
    By assumption, this second quantity is multiplicative, so that
    \begin{align*}
      \opv_p(\Phi \tprod \Psi)
      &\leq \frac{\opv_p(\Phi'_d \tprod \Psi)}{\smallnorm{\nidentity{B}}_p}
      = \frac{\opv_p(\Phi'_d) \opv_p(\Psi)}{\smallnorm{\nidentity{B}}_p}
      \leq \left[ \opv_p(\Phi) + \frac{2}{d^{1/p}} \right] \opv_p(\Psi)
      < \opv_p(\Phi) \opv_p(\Psi) + \epsilon,
    \end{align*}
    where the penultimate inequality is an application of
    Theorem~\ref{thm:approx}.  As epsilon was chosen arbitrarily, the
    multiplicativity of $\opv_p(\Phi'_d)$ for all large enough $d$ implies the
    multiplicativity of $\opv_p(\Phi)$.        
  \end{proof}
\end{theorem}

\section{Minimum Output Entropy and Random Unitary Channels}\label{scn:add}

These results on the multiplicativity of the $p$-norm can be extended
to the additivity of the minimum output entropy.  This is done using
a similar method of proof as the results of the previous
section.
The following theorem demonstrates that the random unitary channel $\Phi'$
constructed in Equation~\eqref{eqn:construction} forms a good approximation
of the original channel $\Phi$, from which the result on the
additivity will follow directly.

\begin{theorem}\label{thm:entropy-approx}
  If $\Phi \in \transform{H,K}$, then the
  random unitary $\Phi' \in \transform{A
    \tprod H, K \tprod B}$ satisfies
  \[ S_{\min}(\Phi) \geq S_{\min}(\Phi') - \log \dm{B} \geq S_{\min}(\Psi) -
  \frac{8 \log \dm{A}}{\dm{A}}. \]

  \begin{proof}    
    Exactly as in Theorem~\ref{thm:approx},
    Proposition~\ref{prop:simulation} implies the first inequality,
    as $\Phi'(\ket 0 \bra 0 \tprod \rho) = \Phi(\rho)
    \tprod \nidentity{B}$.

    Let $\rho$ be a state minimizing $S(\Phi'(\rho))$ and for
    convenience let $\delta = 8 \log \dm{A} / \dm{A}$.
    Equation~\eqref{eqn:output} gives the output of $\Phi'$ on
    $\rho$.  Applying the concavity of the entropy to this, we obtain
    \begin{align*}
      S_{\min}(\Phi')
      = S(\Phi'(\rho))
      \geq q S(\Phi(\sigma) \tprod \nidentity{B}) 
        + (1-q) S(\Phi'(\rho_{S_0^\perp})).
    \end{align*}      
    Applying Corollary~\ref{cor:mixing-entropy} this becomes
    \begin{align*}
      S_{\min}(\Phi')    
      &\geq q S(\Phi(\sigma) \tprod \nidentity{B}) 
        +(1-q)(S(\nidentity{A \tprod H}) - \delta).
    \end{align*}
    Notice that since $\Phi'$ is random unitary, it is the case that
    $\mathcal{A \tprod H}$ is isomorphic to $\mathcal{K \tprod B}$.
    This implies that $S(\nidentity{A \tprod H}) = S(\nidentity{K
      \tprod B})$.  Two additional properties of the entropy will
    be useful:
    $S(\sigma \tprod \xi) = S(\sigma) +
    S(\xi)$ for any $\sigma, \xi$ 
    and $S(\xi) \leq \log \dm{K} = S(\nidentity{K})$ for all $\xi \in
    \density{K}$.  Using these three observations, in order, we find that
    \begin{align*}
      S_{\min}(\Phi')    
      &\geq q S(\Phi(\sigma) \tprod \nidentity{B}) 
        + (1 - q)(S(\nidentity{K \tprod B}) - \delta) \\
      &= q (S(\Phi(\sigma)) + S(\nidentity{B}))
        + (1 - q)(S(\nidentity{K}) + S(\nidentity{B}) - \delta) \\
      &\geq q (S(\Phi(\sigma)) + S(\nidentity{B}))
        + (1 - q)(S(\Phi(\sigma)) + S(\nidentity{B}) - \delta) \\
      &\geq S(\Phi(\sigma)) + S(\nidentity{B}) - \delta.
    \end{align*}
    Finally, since $S(\nidentity{B}) = \log \dm{B}$ and
    $S_{\min}(\Phi) \leq S(\Phi(\xi))$ for any $\xi$, we have
    \begin{align*}
      S_{\min}(\Phi')
      &\geq S_{\min}(\Phi) + \log \dm{B} - \delta,
    \end{align*}
    which completes the proof of the theorem.    
  \end{proof}
\end{theorem}

The proof that the additivity conjecture can be equivalently
restricted to random unitary channels follows from
the previous theorem in a way that is identical to the proof of
Theorem~\ref{thm:pnorm-mult}, with the exception that the $p$-norm has been
replaced by the minimum output entropy.

\begin{theorem}\label{thm:main-result}
  If $\Phi, \Psi \in \transform{H,K}$, then 
  \[ S_{\min}(\Phi \tprod \Psi)  = S_{\min}(\Phi) + S_{\min}(\Psi) \]
  if
  \[ S_{\min}(\Phi'_d \tprod \Psi)  = S_{\min}(\Phi'_d) + S_{\min}(\Psi), \]
  for all sufficiently large $d$, where $\Phi'_d$ is the random unitary
  extension of the channel obtained by applying the construction of
  Section~\ref{scn:channels} to Stinespring dilation for $\Phi$ using
  an ancillary space of dimension $d$.

  \begin{proof}
    Let $\epsilon > 0$, and choose $d$ so that $8 (\log d) / d <
    \epsilon$.  Then, as $\Phi'_d(\ket 0 \bra 0 \tprod \rho) =
    \Phi(\rho) \tprod \nidentity{B}$,
    \[
      S_{\min}(\Phi \tprod \Psi)
      \geq S_{\min}(\Phi'_d \tprod \Psi) - \log \dm{B}.
    \]
    By assumption, this second quantity is additive, so that
    \begin{align*}
      S_{\min}(\Phi \tprod \Psi)
      &\geq S_{\min}(\Phi'_d \tprod \Psi) - \log \dm{B} \\
      &= S_{\min}(\Phi'_d) + S_{\min}(\Psi) - \log \dm{B} \\
      &\geq S_{\min}(\Phi) - \frac{8 \log d}{d} + S_{\min}(\Psi) \\
      &>    S_{\min}(\Phi) + S_{\min}(\Psi) - \epsilon
    \end{align*}
    where the penultimate inequality is an application of
    Theorem~\ref{thm:entropy-approx}.  As $\epsilon$ was chosen
    arbitrarily, the additivity of $\Phi'_d$ for all large enough $d$
    implies the additivity of $\Phi$.
  \end{proof}
\end{theorem}

A direct corollary of this theorem
generalizes a result of Fukuda~\cite{Fukuda07} on the additivity of
the minimum output entropy of unital channels.  This implies
that in the search for either a proof of this conjecture or a
counterexample to it, only random unitary channels need to be considered.

\begin{corollary}
  The additivity of the minimum output entropy, given by
  \[ S_{\min}(\Phi \tprod \Psi) = S_{\min}(\Phi) + S_{\min}(\Psi) \]
  is true for all channels $\Phi$ and $\Psi$ if and only if 
  \[ S_{\min}(\Phi \tprod \Phi) = S_{\min}(\Phi) + S_{\min}(\Phi) \]
  is true for all random unitary channels $\Phi$.

  \begin{proof}
    By a result of Fukuda and Wolf~\cite{FukudaW07} the additivity
    conjecture can be equivalently restricted to the case that $\Psi =
    \Phi$, i.e. that the two channels are the same.  Applying
    Theorem~\ref{thm:main-result} twice results in the statement of
    corollary.
  \end{proof}
\end{corollary}

\section{Circuit Constructions}\label{scn:circuits}

In this section an efficient circuit construction is provided for the
random unitary approximation described in Section~\ref{scn:channels}.
This construction is used to extend the hardness of computationally
distinguishing quantum circuits to the case of random unitary
circuits.

Before constructing these circuits, it will be important to specify
the circuit models that are being used.
The circuit model used to define the quantum circuit
distinguishability problem is the \emph{mixed state quantum circuit} model of
Aharonov, Kitaev, and Nisan~\cite{AharonovK+98}.
Circuits in this model can include unitary gates as well as 
measurements and other non-unitary operations, but as shown
in Ref.~\onlinecite{AharonovK+98}, we may assume that all such circuits first
introduce any necessary ancillary qubits, then perform a unitary
operation, and finally trace out those qubits that are not part 
of the output.  
This approach is equivalent to building a circuit for the
Stinespring dilation of a channel.
As all unitary transformations can be (approximately)
implemented using one and two qubit gates there is no loss in
generality in assuming that the unitary transformations implemented in
such a circuit are composed of gates from some
finite basis of one and two qubit gates.  Circuits in this model can
represent any physically realizable quantum operation.

The second model of quantum circuits we consider is the model of
\emph{random unitary quantum circuits}.  These circuits consist of one
and two qubits gates as well as random unitary gates, which implement
a unitary gate with probability one half.  More
formally, the application of such a gate takes the state $\rho$ to the
state $(1/2) U \rho U^* + (1/2) \rho$,
where $U$ is a one or two qubit unitary gate.
This is an extremely simple model that does not appear to be
universal for the class of transformations that implement random
unitary operations.
It is not clear what is the correct definition of the random unitary
circuit model, and since the aim of the present paper is to prove a
hardness result, an extremely weak definition has been chosen so that
the result will apply to as large a class of circuit models as possible.

One drawback of this weak model is that it is not clear that the
exact construction used in Section~\ref{scn:channels} can be
implemented.  Specifically, the operation $D$ that decoheres the
subspaces $S_0$ and $S_0^\perp$ seems to require a unitary operation
that cannot be decomposed into a series of 
one and two qubit gates, applied with probability one half.
A similar situation occurs for the discrete Weyl operators on the
subspace $S_0^\perp$.  These operations can be implemented in a random
unitary way in a more permissive circuit model, but in order to keep
the hardness result on distinguishing random unitary circuits as
general as possible, a modified construction is presented here.  This
modified construction is built from pieces similar to those used in
Section~\ref{scn:channels}, but the specific building blocks are not
exactly the same.  The construction in this section can also be
applied to the additivity problems, but it is somewhat more
complicated that the construction already presented.

In order to approximate a given circuit with a random unitary circuit
we once again make use of three components.  We once again use
$N,D,$ and $M$ to refer to these components as they play the same
roles as the components used in Section~\ref{scn:channels}, though
they are not exactly the same.
The first two of these
components, $N$ the completely noisy channel and $D$ the complete
dephasing channel, are easy to
implement as random unitary operations in the chosen circuit model.
More difficult to implement is the
channel $M$, which performs a function similar to the channel
described by Equation~\eqref{eqn:mixing-approx}.

The complete dephasing channel $D$ is the channel that sets to zero
all of the off-diagonal elements of a density matrix.  This is
stronger than the operation considered in Section~\ref{scn:channels},
but it is easier to implement as a random unitary circuit.  The
action of this operator applied to the space $\mathcal{A}$, for an
input $\rho$ on $\mathcal{A \tprod H}$ is given by
\begin{equation}\label{eqn:Dcircuit}
  D_{\mathcal{A}}(\rho) 
  = \sum_{i=0}^{\dm{A} - 1} p_i \ket i \bra i \tprod \rho_i,
\end{equation}
where the $p_i$ form a probability distribution.  This operation is
equivalent to measuring the space $\mathcal{A}$ 
in the computational basis and forgetting the result.
The operation $D_{\mathcal{A}}$  can be implemented as a random
unitary circuit by applying the Pauli $Z$ operation to each qubit 
of $\mathcal{A}$
independently with probability 1/2, as described in Ref.~\onlinecite{ChuangY97}.
This will have the effect of negating the off-diagonal elements of a
density matrix with probability 1/2, so that the resulting state is
diagonal in the computational basis.

The completely noisy channel $N$ is also simple to implement as a
random unitary circuit.  This channel can be realized by performing a
uniform mixture of the Pauli operators on each qubit.  This mixture can
be implemented by, independently on each qubit, applying the Pauli $Z$
operation with probability 1/2, followed by applying the Pauli $X$
operation with probability 1/2, as shown in Ref.~\onlinecite{BoykinR03}.  
Intuitively, the $Z$ operations will
zero the off-diagonal elements of a density matrix (viewed in the
computational basis), and the $X$ operations will scramble the
diagonal, resulting in the completely mixed state, $I/2$, on each
qubit.

In Section~\ref{scn:channels} the channel $M$ was implemented as a
completely depolarizing channel on the subspace $S_0^\perp$.  While
the same channel suffices for the circuit case, it is not clear how
this can be implemented using only two-qubit random unitary gates.  To
avoid this difficulty a more complicated construction is used.  This
construction is intuitively the same: it does not affect states in the
subspace $S_0$ of inputs with the $\ket 0$ state in the space
$\mathcal{A}$, and it applies depolarizing noise to states in the
space $S_0^\perp$.  The difference is exactly how this noise is
applied.  The circuit that is constructed will implement the operation
$M$ given by
\begin{equation}\label{eqn:Mcircuit}
   M(\ket i \bra i \tprod \rho) = \begin{cases} 
     \frac{1}{\dim \mathcal{A}}
     (\identity{A} - \ket 0 \bra 0 + \ket{\psi_i} \bra{\psi_i})
     \tprod \nidentity{H} & \text{if $i \neq 0$},\\
     \ket 0 \bra 0 \tprod \rho& \text{if $i=0$},
   \end{cases}
\end{equation}
where $\ket{\psi_i}$ is a nonzero state that depends on $i$, the
exact specification of which will not be significant.

As might be expected, the transformation $M$ can
be implemented using only controlled-mixing operations.  Before
describing this implementation, notice that the controlled application
of the completely depolarizing channel $N$ to a single qubit 
can be described by a random unitary circuit.  
This is because the above implementation of $N$ as a mixture of Pauli $Z$ and
$X$ operations consists only of single qubit gates.
Adding a control
qubit to each of these gates results in two qubit gates, which fit
into the model of random unitary circuits used here.  It is not clear that
general controlled random unitary operations can be implemented as
random unitary circuits in this model, but the
only controlled operation that will be needed for this construction is
the completely depolarizing channel.

\begin{figure}
  \centering
\setlength{\unitlength}{3947sp}%
\begingroup\makeatletter\ifx\SetFigFont\undefined%
\gdef\SetFigFont#1#2#3#4#5{%
  \reset@font\fontsize{#1}{#2pt}%
  \fontfamily{#3}\fontseries{#4}\fontshape{#5}%
  \selectfont}%
\fi\endgroup%
\begin{picture}(2874,2274)(2989,-2623)
\thinlines
{\color[rgb]{0,0,0}\put(3526,-2011){\line( 0,-1){150}}
}%
{\color[rgb]{0,0,0}\put(3751,-2386){\line( 1, 0){2100}}
}%
{\color[rgb]{0,0,0}\put(4951,-1186){\line( 1, 0){150}}
}%
{\color[rgb]{0,0,0}\put(5551,-1186){\line( 1, 0){300}}
}%
{\color[rgb]{0,0,0}\put(5251,-586){\line( 1, 0){600}}
}%
{\color[rgb]{0,0,0}\put(3301,-811){\framebox(450,450){$N$}}
}%
{\color[rgb]{0,0,0}\put(3301,-2011){\framebox(450,450){$N$}}
}%
{\color[rgb]{0,0,0}\put(3901,-1411){\framebox(450,450){$N$}}
}%
{\color[rgb]{0,0,0}\put(4501,-1411){\framebox(450,450){$N$}}
}%
{\color[rgb]{0,0,0}\put(3751,-586){\line( 1, 0){1500}}
}%
{\color[rgb]{0,0,0}\put(4351,-1186){\line( 1, 0){150}}
}%
{\color[rgb]{0,0,0}\put(3751,-1786){\line( 1, 0){2100}}
}%
{\color[rgb]{0,0,0}\put(3526,-1186){\line( 0, 1){375}}
}%
{\color[rgb]{0,0,0}\put(3526,-1186){\line( 0,-1){375}}
}%
{\color[rgb]{0,0,0}\put(4126,-586){\line( 0,-1){375}}
}%
{\color[rgb]{0,0,0}\put(4726,-1786){\line( 0, 1){375}}
}%
{\color[rgb]{0,0,0}\put(3301,-2611){\framebox(450,450){$N$}}
}%
{\color[rgb]{0,0,0}\put(5101,-1411){\framebox(450,450){$N$}}
}%
{\color[rgb]{0,0,0}\put(5326,-2386){\line( 0, 1){975}}
}%
{\color[rgb]{0,0,0}\put(3001,-2386){\line( 1, 0){300}}
}%
{\color[rgb]{0,0,0}\put(3001,-1786){\line( 1, 0){300}}
}%
{\color[rgb]{0,0,0}\put(3001,-1186){\line( 1, 0){900}}
}%
{\color[rgb]{0,0,0}\put(3001,-586){\line( 1, 0){300}}
}%
{\color[rgb]{0,0,0}\put(4126,-586){\circle*{76}}
}%
{\color[rgb]{0,0,0}\put(4726,-1786){\circle*{76}}
}%
{\color[rgb]{0,0,0}\put(5326,-2386){\circle*{76}}
}%
{\color[rgb]{0,0,0}\put(3526,-1186){\circle*{76}}
}%
\end{picture}%
  \caption{One stage of the mixing procedure on the ancillary qubits.
    The mixing operations applied to the qubits in the space
    $\mathcal{H}$ are not shown.}
  \label{fig:mixing}
\end{figure}
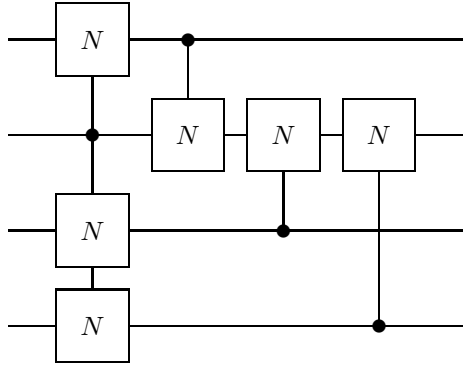

\begin{figure}
  \centering
\setlength{\unitlength}{3247sp}%
\begingroup\makeatletter\ifx\SetFigFont\undefined%
\gdef\SetFigFont#1#2#3#4#5{%
  \reset@font\fontsize{#1}{#2pt}%
  \fontfamily{#3}\fontseries{#4}\fontshape{#5}%
  \selectfont}%
\fi\endgroup%
\begin{picture}(6924,3474)(289,-3673)
\thinlines
{\color[rgb]{0,0,0}\multiput(1201,-3661)(120.61856,0.00000){49}{\line( 1, 0){ 60.309}}
}%
{\color[rgb]{0,0,0}\put(5476,-1786){\line( 0,-1){675}}
}%
{\color[rgb]{0,0,0}\put(5476,-2911){\line( 0,-1){150}}
}%
{\color[rgb]{0,0,0}\put(6076,-586){\line( 0,-1){975}}
}%
{\color[rgb]{0,0,0}\put(6676,-1186){\line( 0,-1){375}}
}%
{\color[rgb]{0,0,0}\multiput(1201,-211)(0.00000,-121.05263){29}{\line( 0,-1){ 60.526}}
}%
{\color[rgb]{0,0,0}\multiput(3151,-211)(0.00000,-121.05263){29}{\line( 0,-1){ 60.526}}
}%
{\color[rgb]{0,0,0}\multiput(5101,-211)(0.00000,-121.05263){29}{\line( 0,-1){ 60.526}}
}%
{\color[rgb]{0,0,0}\multiput(7051,-211)(0.00000,-121.05263){29}{\line( 0,-1){ 60.526}}
}%
{\color[rgb]{0,0,0}\multiput(1201,-211)(120.61856,0.00000){49}{\line( 1, 0){ 60.309}}
}%
{\color[rgb]{0,0,0}\put(601,-2011){\framebox(450,450){$D$}}
}%
{\color[rgb]{0,0,0}\put(601,-1411){\framebox(450,450){$D$}}
}%
{\color[rgb]{0,0,0}\put(601,-811){\framebox(450,450){$D$}}
}%
{\color[rgb]{0,0,0}\put(1351,-1411){\framebox(450,450){$N$}}
}%
{\color[rgb]{0,0,0}\put(1351,-2011){\framebox(450,450){$N$}}
}%
{\color[rgb]{0,0,0}\put(1351,-2911){\framebox(450,450){$N$}}
}%
{\color[rgb]{0,0,0}\put(1351,-3511){\framebox(450,450){$N$}}
}%
{\color[rgb]{0,0,0}\put(1951,-811){\framebox(450,450){$N$}}
}%
{\color[rgb]{0,0,0}\put(2551,-811){\framebox(450,450){$N$}}
}%
{\color[rgb]{0,0,0}\put(3301,-811){\framebox(450,450){$N$}}
}%
{\color[rgb]{0,0,0}\put(3301,-2011){\framebox(450,450){$N$}}
}%
{\color[rgb]{0,0,0}\put(3901,-1411){\framebox(450,450){$N$}}
}%
{\color[rgb]{0,0,0}\put(4501,-1411){\framebox(450,450){$N$}}
}%
{\color[rgb]{0,0,0}\put(5251,-1411){\framebox(450,450){$N$}}
}%
{\color[rgb]{0,0,0}\put(5251,-811){\framebox(450,450){$N$}}
}%
{\color[rgb]{0,0,0}\put(5851,-2011){\framebox(450,450){$N$}}
}%
{\color[rgb]{0,0,0}\put(6451,-2011){\framebox(450,450){$N$}}
}%
{\color[rgb]{0,0,0}\put(3301,-2911){\framebox(450,450){$N$}}
}%
{\color[rgb]{0,0,0}\put(3301,-3511){\framebox(450,450){$N$}}
}%
{\color[rgb]{0,0,0}\put(5251,-2911){\framebox(450,450){$N$}}
}%
{\color[rgb]{0,0,0}\put(5251,-3511){\framebox(450,450){$N$}}
}%
{\color[rgb]{0,0,0}\put(301,-586){\line( 1, 0){300}}
}%
{\color[rgb]{0,0,0}\put(1051,-586){\line( 1, 0){900}}
}%
{\color[rgb]{0,0,0}\put(2401,-586){\line( 1, 0){150}}
}%
{\color[rgb]{0,0,0}\put(3001,-586){\line( 1, 0){300}}
}%
{\color[rgb]{0,0,0}\put(3751,-586){\line( 1, 0){1500}}
}%
{\color[rgb]{0,0,0}\put(5701,-586){\line( 1, 0){1500}}
}%
{\color[rgb]{0,0,0}\put(301,-1186){\line( 1, 0){300}}
}%
{\color[rgb]{0,0,0}\put(301,-1786){\line( 1, 0){300}}
}%
{\color[rgb]{0,0,0}\put(1051,-1786){\line( 1, 0){300}}
}%
{\color[rgb]{0,0,0}\put(1051,-1186){\line( 1, 0){300}}
}%
{\color[rgb]{0,0,0}\put(4951,-1186){\line( 1, 0){300}}
}%
{\color[rgb]{0,0,0}\put(4351,-1186){\line( 1, 0){150}}
}%
{\color[rgb]{0,0,0}\put(6301,-1786){\line( 1, 0){150}}
}%
{\color[rgb]{0,0,0}\put(1801,-1186){\line( 1, 0){2100}}
}%
{\color[rgb]{0,0,0}\put(1801,-1786){\line( 1, 0){1500}}
}%
{\color[rgb]{0,0,0}\put(3751,-1786){\line( 1, 0){2100}}
}%
{\color[rgb]{0,0,0}\put(6901,-1786){\line( 1, 0){300}}
}%
{\color[rgb]{0,0,0}\put(5701,-1186){\line( 1, 0){1500}}
}%
{\color[rgb]{0,0,0}\put(301,-2686){\line( 1, 0){1050}}
}%
{\color[rgb]{0,0,0}\put(1801,-2686){\line( 1, 0){1500}}
}%
{\color[rgb]{0,0,0}\put(3751,-2686){\line( 1, 0){1500}}
}%
{\color[rgb]{0,0,0}\put(5701,-2686){\line( 1, 0){1500}}
}%
{\color[rgb]{0,0,0}\put(301,-3286){\line( 1, 0){1050}}
}%
{\color[rgb]{0,0,0}\put(1801,-3286){\line( 1, 0){1500}}
}%
{\color[rgb]{0,0,0}\put(5701,-3286){\line( 1, 0){1500}}
}%
{\color[rgb]{0,0,0}\put(3751,-3286){\line( 1, 0){1500}}
}%
{\color[rgb]{0,0,0}\put(1576,-586){\line( 0,-1){375}}
}%
{\color[rgb]{0,0,0}\put(1576,-1411){\line( 0,-1){150}}
}%
{\color[rgb]{0,0,0}\put(1576,-2011){\line( 0,-1){450}}
}%
{\color[rgb]{0,0,0}\put(1576,-2911){\line( 0,-1){150}}
}%
{\color[rgb]{0,0,0}\put(2176,-1186){\line( 0, 1){375}}
}%
{\color[rgb]{0,0,0}\put(2776,-1786){\line( 0, 1){975}}
}%
{\color[rgb]{0,0,0}\put(3526,-1186){\line( 0, 1){375}}
}%
{\color[rgb]{0,0,0}\put(3526,-2011){\line( 0,-1){450}}
}%
{\color[rgb]{0,0,0}\put(3526,-2911){\line( 0,-1){150}}
}%
{\color[rgb]{0,0,0}\put(3526,-1186){\line( 0,-1){375}}
}%
{\color[rgb]{0,0,0}\put(4126,-586){\line( 0,-1){375}}
}%
{\color[rgb]{0,0,0}\put(4726,-1786){\line( 0, 1){375}}
}%
{\color[rgb]{0,0,0}\put(5476,-1786){\line( 0, 1){375}}
}%
{\color[rgb]{0,0,0}\put(5476,-961){\line( 0, 1){150}}
}%
{\color[rgb]{0,0,0}\put(1576,-586){\circle*{76}}
}%
{\color[rgb]{0,0,0}\put(5476,-1786){\circle*{76}}
}%
{\color[rgb]{0,0,0}\put(2176,-1186){\circle*{76}}
}%
{\color[rgb]{0,0,0}\put(2776,-1786){\circle*{76}}
}%
{\color[rgb]{0,0,0}\put(4126,-586){\circle*{76}}
}%
{\color[rgb]{0,0,0}\put(4726,-1786){\circle*{76}}
}%
{\color[rgb]{0,0,0}\put(6076,-586){\circle*{76}}
}%
{\color[rgb]{0,0,0}\put(6676,-1186){\circle*{76}}
}%
{\color[rgb]{0,0,0}\put(3526,-1186){\circle*{76}}
}%
\end{picture}%
  \caption{Circuit performing the conditional mixing procedure $M
    \circ D$.  The top three qubits are simulating the ancillary
    qubits of the original circuit in the space $\mathcal{A}$, and the
    bottom two are simulating the input to the original circuit in the
    space $\mathcal{H}$.  The dashed lines separate the stages of the
    mixing procedure.}
  \label{fig:ancilla}
\end{figure}
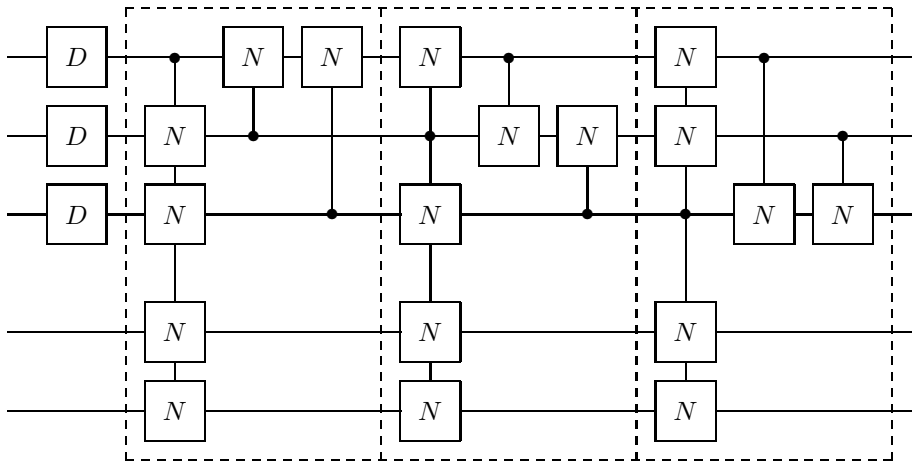

Let $m$ be the number of qubits in the space $\mathcal{A}$ that are
given as part of the input to $M$, i.e. the number of ancillary qubits
used to represent the ancillary space used by the original channel.
The implementation of $M$ consists of $m$ stages, with the
$j$th stage testing that the $j$th qubit of the space $\mathcal{A}$ is
in the $\ket 0$ state, and mixing the qubits if this is not the case.
An example of one stage of the circuit
is given in Figure~\ref{fig:mixing}.
The $j$th stage consists first of an
application of the controlled $N$ operation from the $j$th qubit to
each other qubit of $\mathcal{A \tprod H}$.
After these operations, stage $j$ is completed by $m-1$ further
controlled $N$ operations: each with the $j$th qubit as the target
qubit and one of the other qubits of $\mathcal{A}$ as the control qubit.  An
example of this construction with $m=3$ is presented in
Figure~\ref{fig:ancilla}.

Given these circuit implementations of the three channels $D, N, M$, the
random unitary circuit $C$ that approximates a given circuit $Q$ is
constructed in exactly the same was as in
Equation~\eqref{eqn:construction}.  More concretely, let $Q$ be a
circuit implementing the operation
\[ Q(\rho) = \ptr{B} U (\ket 0 \bra 0 \tprod \rho) U^*, \]
where the ancillary qubits are in the space $\mathcal{A}$.  The
circuit $C$ that approximates it is then given by
\begin{equation}\label{eqn:circuit_construction}
  C(\rho) =   
  N_\mathcal{B} \left( 
    U \left[ 
      (M \circ D_\mathcal{A})(\rho) 
    \right] U^* 
  \right).
\end{equation}
This circuit $C$ is given by a random unitary circuit, since it is
the composition of smaller random unitary circuits.  As the operations
$D_{\mathcal{A}}$ and $M$ do not affect inputs of the form $\ket 0
\bra 0 \tprod \rho$, the proof of Proposition~\ref{prop:simulation}
holds also for the circuit case, so that
\begin{equation}\label{eqn:circuit_simulation}
  C(\ket 0 \bra 0 \tprod \sigma) = Q(\sigma) \tprod \nidentity{B}.
\end{equation}
Combining this with equation~\eqref{eqn:Dcircuit} and the fact that applying
$D_\mathcal{A}$ twice has no further effect, the output of
$C$ on an arbitrary input state $\rho$ is of the form
\begin{equation}\label{eqn:circuit_output}
  C(\rho)
  = \sum_{i=0}^{\dm{A} - 1} p_i C(\ket i \bra i \tprod \rho_i)
  = p_0 Q(\rho_0) \tprod \nidentity{B} +\sum_{i=1}^{\dm A - 1}
        p_i C(\ket i \bra i \tprod \rho_i).
\end{equation}
In the remainder of the paper it
is shown that this construction does not significantly alter the
distinguishability properties of quantum circuits.

As a first step towards this, it is shown that the above circuit
construction correctly implements the channel $M$ described by Equation~\ref{eqn:Mcircuit}.  Much of the proof
of this lemma is similar to the proof of
Lemma~\ref{lem:ancilla-mixing}, but the operation $M$ considered in
this section is slightly different and we also need to extend the
lemma to the case that there is an additional reference system.  This
system, given by the space $\mathcal{F}$, is needed in the case of
distinguishability, as a party attempting to distinguish two channels
is permitted to use a portion of a larger entangled state as input to
the channels.  This can be seen from the definition of the
distinguishability problem, which is given in the next section.

\begin{lemma}\label{lem:ancilla-mixing-aux}
  On input states of the form $\ket k \bra k \tprod \rho \in
  \density{A \tprod H \tprod F}$ for $\ket k \bra k \in \density{A}$
  with $0 < k \leq 2^m-1$, the output of $C$ satisfies
  \[ 
     \tnorm{(C \tprod \identity{F})(\ket k \bra k \tprod \rho) - 
       \nidentity{A \tprod H}
         \tprod \ptr{H} \rho}
     \leq \frac{1}{2^{m-1}},\]
     where $m$ is the number of ancillary qubits used by the circuit $Q$.

  \begin{proof}
    On input of the form $\ket k \bra k \tprod \rho$ the decoherence
    operations that are applied to the qubits in $\mathcal{A}$ can be
    ignored, as they have no effect on qubits in a state of the
    computational basis.  As $k \neq 0$ at least one qubit is in the
    state $\ket 1$, and so the controlled mixing operations in the
    implementation of the channel $M$ will have
    an effect.  Let the first nonzero qubit among the qubits of
    $\mathcal{A}$ be the $j$th one.
    The first controlled $N$ operation with nonzero control qubit
    that effects the $j$th qubit will be at the $j$th stage of the
    mixing process, where the $j$th qubit is the control qubit.  As
    this qubit is not modified before this stage (as any previous
    qubits are in the state $\ket 0$ by choice of $j$), the first
    $m-1$ gates
    in the $j$th stage will mix the remaining qubits, so that the
    state after these gates is, using Equation~\eqref{eqn:traceout},
    \[ \ket 1 \bra 1 \tprod \nidentity{A'} \tprod \nidentity{H} \tprod
    \ptr{H} \rho, \]
    where for notational
    convenience the $j$th qubit has been written first, and
    $\mathcal{A'}$ is the space of all but the $j$th qubit of
    $\mathcal{A}$.
    The remainder of the $j$th stage of the mixing process consists of
    $m-1$ controlled $N$ gates with the $j$th qubit as the
    target, each controlled by one of the $m-1$ qubits in
    $\mathcal{A'}$.  Considering the state $I/2^{m-1}$ on $\mathcal{A'}$ in
    the computational basis, the only term for which qubit $j$ is not
    mixed by these operations is the all zero term.  With this
    observation, the state after the $j$th stage is
    \begin{align*}
      \frac{1}{2^{m-1}} & \left[ 
        \ket{1}\bra{1} \tprod (\ket 0 \bra 0)^{\tprod m-1}
        + \frac{\ket 0 \bra 0 + \ket 1 \bra 1}{2} \tprod
        (\identity{A'} -  (\ket 0 \bra 0)^{\tprod m-1})
        \right] 
        \tprod \nidentity{H}
        \tprod \ptr{H} \rho \nonumber \\
      &= \frac{\identity{A} 
           + \ket{1}\bra{1} \tprod (\ket 0 \bra 0)^{\tprod m-1}
           - (\ket 0 \bra 0)^{\tprod m}}{2^m}
         \tprod \nidentity{H}
         \tprod \ptr{H} \rho.
    \end{align*}
    This proves that the circuit implementing the channel $M$ does so
    correctly, as this quantity is exactly
    the state given in Equation~\eqref{eqn:circuit_output}
    with the addition of $\ptr{H} \rho$ in the reference system.

    As in the proof of Lemma~\ref{lem:ancilla-mixing}, let this state
    be $\sigma$.  Computing the distance from this state to the
    desired one, we have
    \begin{equation*}
      \tnorm{ \sigma - \nidentity{A} \tprod \nidentity{H}
         \tprod \ptr{H} \rho}
      = \frac{1}{2^m} 
         \tnorm{\ket{1}\bra{1} \tprod (\ket 0 \bra 0)^{\tprod m-1} 
           - (\ket 0 \bra 0)^{\tprod m}}
      = \frac{1}{2^{m-1}}.
    \end{equation*}
    Finally, by noting that the remainder of the circuit $C$ is
    random unitary, an application of Lemma~\ref{lem:ru-dist-noise}
    yields the desired bound.    
  \end{proof}
\end{lemma}

\section{\class{QIP}-Completeness of Distinguishing Random Unitary
  Circuits}\label{scn:distinguish}

The construction outlined in the previous section can be used to 
show that the problem of distinguishing random unitary
quantum circuits is \class{QIP}-complete, where \class{QIP} is the
class of problems having quantum interactive proof systems.
The basic idea is to reduce
an instance of the quantum circuit distinguishability problem to one
with random unitary circuits that has the same distinguishability properties.
This will be done by taking the instance $(Q_1, Q_2)$ and constructing
the instance $(C_1, C_2)$ by applying the construction of
Section~\ref{scn:circuits} to each of these circuits.  The quantum
circuit distinguishability problem is given by
\begin{probenv}{Quantum Circuit Distinguishability}
  For constants $0 \leq b < a \leq 2$, the input consists of quantum
  circuits $Q_1$ and $Q_2$ that implement transformations from
  $\mathcal{H}$ to $\mathcal{K}$.
  The promise problem is to distinguish the two cases:
  \begin{description}
    \item[Yes] $\dnorm{Q_1 - Q_2} \geq a$,
    \item[No] $\dnorm{Q_1 - Q_2} \leq b$.
  \end{description}
\end{probenv}
This problem was introduced and shown to be complete for the
complexity class \class{QIP} in Ref.~\onlinecite{RosgenW05}.
The norm used in the definition of the problem is the diamond norm,
which can be defined on a channel $\Phi \in \transform{H,K}$ by
\[ \dnorm{\Phi} = \sup_{\tnorm{X} = 1} \tnorm{(\Phi \tprod
  \identity{F})(X)}, \]
where the space $\mathcal{F}$ has dimension at least as large as
$\mathcal{H}$.  A more thorough definition as well as some properties
of this norm can be found in Ref.~\onlinecite{KitaevS+02}.  
It may be helpful to note that the diamond norm of $\Phi$ is just the
completely bonded norm of the adjoint channel $\Phi^*$, where the
adjoint is taken with
respect to the Hilbert-Schmidt inner product.
It is shown in Ref.~\onlinecite{RosgenW05} that the maximum of this
norm on the difference of two completely positive transformations is
achieved by a density matrix, and so we can restrict the supremum in
the definition to $\density{H \tprod F}$.

Here we consider this distinguishability problem with the added
restriction that the input circuits are random unitary circuits in the
model defined in Section~\ref{scn:circuits}.
The following theorem states that the constructed circuits
$C_1$ and $C_2$ have almost the same distinguishability
characteristics as the original circuits $Q_1$ and $Q_2$.  As the
circuit distinguishability problem is defined as a promise problem,
this theorem shows immediately that the problem of distinguishing
random unitary circuits is \class{QIP}-complete, as the construction
of the circuits $C_1$ and $C_2$ can be performed efficiently.

\begin{theorem}\label{thm:reduction}
  For any $\epsilon > 0$,
  \[ \dnorm{Q_1 - Q_2}
     \leq \dnorm{C_1 - C_2} 
     \leq \dnorm{Q_1 - Q_2} + \epsilon, \]
  where the circuits $C_1$ and $C_2$ use $O(\log 1/\epsilon)$
  ancillary qubits.

  \begin{proof}
    The first inequality is not hard to show.  Once again, if the
    state $(\ket 0 \bra 0)^{\tprod m} \tprod \rho$ is given as input
    to the circuit $C_i$, then by Equation~\ref{eqn:circuit_simulation},
    the output is a simulation of $Q_i$, so that the
    distinguishability of $Q_1 $ and $Q_2$ cannot be greater than the
    distinguishability of $C_1$ and $C_2$.
    More formally, note that
    \begin{align*}
      \dnorm{Q_1 - Q_2}
      = \sup_{\rho \in \density{H \tprod F}}
          \tnorm{(Q_1 \tprod \identity{F})(\rho) 
                -(Q_2 \tprod \identity{F})(\rho)},
    \end{align*}
    and fix $\delta>0$ and $\rho$ as a state achieving a value within
    $\delta$ of this supremum.  By Equation~\ref{eqn:circuit_simulation},
    if the state
    $(\ket 0 \bra 0)^{\tprod m} \tprod \rho$ is given as input to
    the circuit $C_i$, then the output is given by
    $(Q_i \tprod \identity{F}) (\rho)$.
    Using this we have
    \begin{align*} 
      \dnorm{C_1 - C_2}
      & \geq \tnorm{ 
        (C_1 \tprod \identity{F})
          ((\ket 0 \bra 0)^{\tprod m} \tprod\rho)
        - (C_2 \tprod \identity{F})
          ((\ket 0 \bra 0)^{\tprod m} \tprod\rho)} \\
      &= \tnorm{ (Q_1 \tprod \identity{F}) (\rho)
        - (Q_2 \tprod \identity{F}) (\rho)} \\
      &\geq \dnorm{Q_1 - Q_2} - \delta.
    \end{align*}
    Since this is true for any $\delta > 0$, it must be the case that
    $\dnorm{Q_1 - Q_2} \leq \dnorm{C_1 - C_2}$.

    The second inequality requires somewhat more work.  Let $m$ be the
    number of ancillary qubits and let $n$
    be the number of input qubits used by the circuits $Q_i$, so that
    $m = \ceil{\log \dm{A}}$ and $n = \ceil{\log \dm{H}}$.  Without
    loss of generality let $2^{-(m-3)} < \epsilon$, 
    by adding at most $3 + \log (1/\epsilon)$ extra (unused) ancillary
    qubits to $Q_1$ and $Q_2$.
    Let $\rho \in \density{A \tprod H \tprod F}$ be a state such that
    \[ \dnorm{C_1 - C_2} - \epsilon/2
       \leq \tnorm{(C_1 \tprod \identity{F})(\rho)
               -(C_2 \tprod \identity{F})(\rho)}, \]
    and note that the reference system $\mathcal{F}$ need not have the same
    dimension as the space of the same name considered in the proof of
    the previous inequality.  The first gates applied in the circuit
    $C_i$ are the decoherence gates applied to $\mathcal{A}$.
    These gates produce a state of the form
    $ \sum_{i=0}^{2^m - 1} p_i \ket i \bra i \tprod
       \sigma_i, $
    and since applying these gates twice has no further effect, the output of
    the circuits $C_1$ and $C_2$ is the same on $\rho$ as it is on
    this state.  Applying the triangle inequality, the quantity of
    interest is
    \begin{equation}\label{eqn:soundness1}
      \dnorm{C_1 - C_2} - \epsilon/2
      \leq \sum_{i=0}^{2^m - 1} p_i
        \tnorm{ (C_1 \tprod \identity{F})(\ket i \bra i \tprod\sigma_i)
              - (C_2 \tprod \identity{F})(\ket i \bra i
      \tprod\sigma_i)}
    \end{equation}
    Then, by applying Lemma~\ref{lem:ancilla-mixing-aux} to each term
    with $i \neq 0$ the states in the norm can be replaced with completely
    mixed states on $\mathcal{A \tprod H}$ plus a small correction
    factor.  Doing this for each of these terms we have
    \begin{align*}
      p_i &\tnorm{ 
                (C_1 \tprod \identity{F})(\ket i \bra i \tprod\sigma_i)
              - (C_2 \tprod \identity{F})(\ket i \bra i \tprod\sigma_i)}\\
      &\leq p_i \left[ \frac{2}{2^{m - 1}} +
          \tnorm{
              \nidentity{A \tprod H} \tprod \ptr{H} \sigma_i
            - \nidentity{A \tprod H} \tprod \ptr{H} \sigma_i}
          \right] \\
      &= p_i / 2^{m-2} < p_i \epsilon/2.
    \end{align*}
    Applying this to Equation~\eqref{eqn:soundness1} we have
    \begin{equation*} 
      \dnorm{C_1 - C_2} - \epsilon/2
      \leq 
      p_0 \tnorm{ (C_1 \tprod \identity{F})(\ket 0 \bra 0 \tprod\sigma_0)
        - (C_2 \tprod \identity{F})(\ket 0 \bra 0 \tprod\sigma_0)}
      + \sum_{i=1}^{2^m - 1} p_i \epsilon/2.
    \end{equation*}
    By Equation~\ref{eqn:circuit_simulation} the output of the circuit
    $C_i$ on this input can be replaced the output of the circuit
    $Q_i$ and a maximally mixed state.
    When this is done to the previous equation,
    the desired bound is given by
    \begin{align*}
       \dnorm{C_1 - C_2}
       & \leq p_0 \tnorm{ (Q_1 \tprod \identity{F})(\sigma_0)
              - (Q_2 \tprod \identity{F})(\sigma_0)}
       + (1 - p_0) \epsilon/2  + \epsilon/2
        \leq \dnorm{Q_1 - Q_2} + \epsilon.
    \end{align*}
    This completes the proof of the theorem, as $0 \leq p_0 \leq 1$.    
  \end{proof}
\end{theorem}

\section{Conclusion}

A method for approximating a quantum channel with one that is random
unitary has been provided.  This approximation yields the equivalence
of several important problems when restricted to random unitary
channels.  These results raise the open problem of how far these
equivalences extend.  What other problems can be restricted to the
random unitary case without loss of generality, and what problems
are simplified when restricted to this class of channels?

\section*{Acknowledgements}

I would like to thank John Watrous for several helpful discussions, as
well as Michael Wolf and the anonymous referees for
their comments, including a suggestion by one referee that simplified
the argument in Section~\ref{scn:channels}.
Canada's NSERC and MITACS have supported this research.


\end{document}